\documentstyle[sprocl]{article}

\def\be{\beta}
\def\ga{\gamma}
\def\de{\delta}
\def\ep{\epsilon}

\def\ka{\kappa}
\def\la{\lambda}

\def\si{\sigma}

\def\ph{\phi}

\def\ps{\psi}

\def\De{\Delta}

\def\cl{{\cal L}}

\def\frac#1#2{{\textstyle{{#1}\over {#2}}}}
\def\half{{\textstyle{1\over 2}}}
\def\lsim{\mathrel{\rlap{\lower4pt\hbox{\hskip1pt$\sim$}}
    \raise1pt\hbox{$<$}}}
\def\gsim{\mathrel{\rlap{\lower4pt\hbox{\hskip1pt$\sim$}}
    \raise1pt\hbox{$>$}}}
\def\sqr#1#2{{\vcenter{\vbox{\hrule height.#2pt
         \hbox{\vrule width.#2pt height#1pt \kern#1pt
         \vrule width.#2pt}
         \hrule height.#2pt}}}}

\def\lrprtmu{\stackrel{\leftrightarrow}{\partial_\mu}}
\def\lrprtnu{\stackrel{\leftrightarrow}{\partial^\nu}}

\def\Im{\hbox{Im}\,}

\newcommand{\beq}{\begin{equation}}
\newcommand{\eeq}{\end{equation}}
\newcommand{\bea}{\begin{eqnarray}}
\newcommand{\eea}{\end{eqnarray}}
\newcommand{\rf}[1]{(\ref{#1})}

\begin{document}

\begin{flushright}
IUHET 387\\
April 1998
\end{flushright}
\vglue 0.5 cm

\title{CPT AND LORENTZ TESTS IN THE STANDARD MODEL\footnote
{Presented at PASCOS 98, 
Northeastern University, Boston, March 1998} 
}

\author{V.\ ALAN KOSTELECK\'Y}

\address{Physics Department, Indiana University\\
Bloomington, IN 47405, U.S.A.\\
Email: kostelec@indiana.edu} 

\maketitle\abstracts{ 
A general Lorentz-violating extension of the minimal 
SU(3) $\times$ SU(2)$ \times$ U(1) standard model
is described,
including both CPT-even and CPT-odd terms.
Some theoretical properties and experimental implications are given.
The extension can be regarded as
a low-energy limit of a physically relevant 
Lorentz-covariant fundamental theory 
in which spontaneous Lorentz violation occurs.
}

\section{Introduction}

The minimal SU(3)$\times$SU(2)$\times$U(1) standard model
provides an accurate phenomenological description of nature
at energies low compared to the Planck scale $M_P$.
The model is presumably the low-energy limit of a 
fundamental theory that consistently combines gravity
and quantum mechanics. 
Experiments with present technology 
searching for effects from this fundamental theory 
must contend with the difference between the 
electroweak scale $m_W$ and $M_P$,
which is about 17 orders of magnitude
and is likely to induce a strong suppression of possible signals.
This makes it of interest to identify 
experiments that are of exceptional sensitivity 
and are seeking effects excluded 
in conventional renormalizable gauge theories.
One approach is to look for qualitatively new physics  
in candidate fundamental theories.
For example,
in string (M) theory
there are qualitatively new effects 
expected at the Planck scale,
and one can consider the possibility 
of associated low-energy signals. 

In this talk,
I examine the idea 
that the new physics involves 
a spontaneous violation of Lorentz symmetry.\cite{kps}
This occurs in theories with Lorentz-covariant dynamics
that exhibit field interactions destabilizing the naive vacuum 
and inducing nonzero expectation values for Lorentz tensors.
Certain string theories may be of this type.
If the nonzero tensor expectation values
are oriented in the physical four spacetime dimensions,
apparent Lorentz violations could occur 
at the level of the standard model.

The Lorentz transformations
are linked to the discrete transformations
of charge conjugation C,
parity reflection P,
and time reversal T through the CPT theorem.\cite{sachs}
Under mild assumptions,
this theorem implies that CPT must be a symmetry
of local relativistic quantum field theories.
It follows that violations of both Lorentz and CPT symmetry
would represent potentially observable signals
arising at the level of the fundamental theory
and lying outside conventional renormalizable gauge theory.
With the expected severe suppression,
only specialized experiments would be sensitive 
to such effects.

\section{Standard-Model Extension and QED Limit}

The effects of spontaneous Lorentz and CPT breaking  
at the level of the standard model
can be treated by including certain terms
that appear to violate these symmetries.
A general Lorentz-violating extension of the minimal 
SU(3) $\times$ SU(2)$ \times$ U(1) standard model
that includes both CPT-even and CPT-odd terms 
exists.\cite{cksm}
The usual gauge invariances and gauge symmetry breaking are valid
and the theory is hermitian and power-counting renormalizable.
At present,
this standard-model extension seems to be the only candidate
for a consistent extension of the standard model
providing a microscopic theory of Lorentz violation.
It must be the low-energy limit
of any fundamental theory (not necessarily string theory) 
that generates the standard model
and contains spontaneous Lorentz and CPT violation.

Although Lorentz symmetry is spontaneously broken,
the standard-model extension retains
many attractive features of Lorentz-covariant theories.\cite{cksm}
One reason is that
Lorentz symmetry holds in the underlying fundamental theory,
so properties such as microcausality 
and positivity of the energy
are expected to hold. 
Standard quantization methods are unaffected,
and energy and momentum are conserved 
if the tensor expectation values arising from the symmetry breaking 
are independent of spacetime position.
Also,
the standard-model extension remains invariant under
rotations or boosts of the observer's inertial frame
(\it observer \rm Lorentz transformations).
Lorentz violations arise only through the 
existence of nonzero tensor expectation values in the vacuum,
which are revealed by
rotations or boosts of the (localized) fields only
(\it particle \rm Lorentz transformations).

The usual standard model disregards gravitational effects,
and the same is true of the Lorentz-violating extension.
This means the broken particle Lorentz symmetry 
appears as a global symmetry,
so Nambu-Goldstone modes might be expected.
The effect of these modes would be of interest
if gravity were included,
since local Lorentz invariance would be involved.
In gauge theories,
if global symmetries are promoted to local ones
then the Higgs mechanism can occur,
modifying the propagator of the gauge fields
to generate a mass and eliminating the Nambu-Goldstone modes. 
However,
in a theory with gravity
there is no gravitational Higgs effect:
the graviton propagator can be modified
if a Lorentz tensor acquires a nonzero expectation value,
but no mass for the graviton arises 
because the gravitational connection 
involves derivatives of the metric.\cite{kps}

The standard-model extension may be constructed
by adding all possible Lorentz-violating terms
that are compatible with spontaneous symmetry breaking
in an underlying theory and that preserve
SU(3) $\times$ SU(2)$ \times$ U(1) gauge invariance
and power-counting renormalizability.
The explicit form of the extension,
including both CPT-even and CPT-odd terms,
can be found in the literature.\cite{cksm}

The existence of numerous sensitive experimental tests
of CPT and Lorentz invariance 
in the context of quantum electrodynamics (QED)
makes it of value to consider associated
limiting cases of the standard-model extension.
These limits are generalizations of QED
that include modifications in
both the fermion and the photon sectors
allowing for possible Lorentz and CPT violations.\cite{cksm}
For the special case of the extended
theory of photons, electrons, and positrons,
the associated lagrangian has a relatively simple form.
The usual terms are:
\beq
\cl^{\rm QED} =
\overline{\ps} \ga^\mu (\half i \lrprtmu - q A_\mu ) \ps 
- m \overline{\ps} \ps 
- \frac 1 4 F_{\mu\nu}F^{\mu\nu}
\quad .
\label{a}
\eeq
There are two CPT-violating terms in the fermion sector
and one in the photon sector:
\beq
\cl^{\rm CPT}_{e} =
- a_{\mu} \overline{\ps} \ga^{\mu} \ps 
- b_{\mu} \overline{\ps} \ga_5 \ga^{\mu} \ps \quad ,
$$
$$
\cl^{\rm CPT}_{\ga} =
\half (k_{AF})^\ka \ep_{\ka\la\mu\nu} A^\la F^{\mu\nu}
\quad .
\label{b}
\eeq
Certain Lorentz-violating but CPT-preserving terms are
also possible:
\beq
\cl^{\rm Lorentz}_{e} = 
c_{\mu\nu} \overline{\ps} \ga^{\mu} 
(\half i \lrprtnu - q A^\nu ) \ps 
+ d_{\mu\nu} \overline{\ps} \ga_5 \ga^\mu 
(\half i \lrprtnu - q A^\nu ) \ps 
- \half H_{\mu\nu} \overline{\ps} \si^{\mu\nu} \ps 
$$
$$
\cl^{\rm Lorentz}_{\ga} =
-\frac 1 4 (k_F)_{\ka\la\mu\nu} F^{\ka\la}F^{\mu\nu}
\quad .
\label{c}
\eeq
The various coefficients appearing in the above terms
play the role of Lorentz- and CPT-violating couplings
and are expected to be suppressed by many orders of magnitude.
Information about the notation used above can be found
in the literature,\cite{cksm}
along with more details about related issues.
For example,
field redefinitions can be used to show that certain components 
of the coupling coefficients are physically unobservable.

\section{Experimental Tests}

Although many tests of Lorentz invariance and CPT exist,
most are insensitive to effects in the standard-model extension 
because the coupling coefficients are small.
However,
some experiments provide interesting constraints.
The quantitative microscopic theory of Lorentz and CPT violation
offered by the standard-model extension
can be used 
both to analyze and compare bounds obtained from various experiments
and to identify potential experimental signals.
To date,
investigations along these lines have been performed
for possible bounds on CPT and Lorentz violations 
from measurements of 
neutral-meson oscillations,\cite{kexpt,ckpv,bexpt,ak}
from tests of QED 
in Penning traps,\cite{pennexpts,bkr}
from hydrogen and antihydrogen spectroscopy,\cite{antih,bkr2}
from photon properties,\cite{cksm}
and from baryogenesis.\cite{bckp}
In what follows,
a brief outline of a few of these results is presented.
Several other investigations are underway,
including a study\cite{kla}
of limits attainable in
clock-comparison experiments.\cite{cc}

\subsection{Tests with Neutral-Meson Oscillations}

First,
consider interferometry with neutral
$K$, $D$, $B_d$, or $B_s$ mesons.
Two types of (indirect) CP violation
can be explored:
T violation with CPT invariance,
or CPT violation with T invariance.
These can be described phenomenologically using 
complex parameters $\ep_P$ and $\de_P$,
respectively,
where the subscript $P$ labels the type of neutral meson.
The effective hamiltonian for the time evolution 
of a $P$-meson state incorporates $\ep_P$ and $\de_P$,
so experiments investigating oscillations
can provide some bounds on CPT violation.

An expression for the CPT-violating parameter $\de_P$ 
can be derived in the context 
of the standard-model extension.\cite{ak}
It depends on only one type of coupling coefficient,
arising from CPT-violating terms of the form
$- a^q_{\mu} \overline{q} \ga^\mu q$,
where $q$ is a quark field
and the coefficient $a^q_{\mu}$ is spacetime constant 
but varies with quark flavor $q$.
The sensitivity of neutral-meson systems 
to $a^q_{\mu}$ is a consequence of the flavor-changing oscillations.
Since none of the other experiments mentioned in this talk 
involve flavor-changing effects,
none can provide experimental signals for coupling coefficients
of the type $a^q_{\mu}$.

Potentially important effects arise because 
the CPT and Lorentz breaking
generates a dependence of $\de_P$ 
on the boost and orientation of the meson.\cite{ak}
Denote the $P$-meson four-velocity by
$\be^\mu \equiv \ga(1,\vec\be)$.
Then,
the standard-model extension implies that the expression
\beq
\de_P \approx i \sin\hat\ph \exp(i\hat\ph) 
\ga(\De a_0 - \vec \be \cdot \De \vec a) /\De m
\quad 
\label{e}
\eeq
holds at leading order in all coupling coefficients.
Here,
$\hat\ph\equiv \tan^{-1}(2\De m/\De\ga)$,
where $\De m$ and $\De \ga$
are the mass and decay-rate differences
between the $P$-meson eigenstates,
respectively,
and $\De a_\mu \equiv a_\mu^{q_2} - a_\mu^{q_1}$,
where $q_1$ and $q_2$ represent the valence-quark flavors 
in the $P$ meson.
For simplicity,
various subscripts $P$ have been omitted from 
variables on the right-hand side of Eq.\ \rf{e}.

Several consequences for experiment are implied by 
the form of Eq.\ \rf{e}.
For one,
the real and imaginary parts of $\de_P$ are proportional.\cite{ckpv}
For another,
the magnitude of $\de_P$ can vary with $P$ 
because the couplings $a_\mu^q$ are flavor dependent.\cite{ckpv}
Still others arise from 
the dependence of $\de_P$ on
the boost magnitude and orientation,
suggesting a variety of possibilities for signals
of Lorentz and CPT breaking.\cite{ak}
For instance,
boosted mesons may exhibit larger CPT-violating effects,
so otherwise comparable experiments seeking to bound $\de_P$ 
might have different CPT reaches.

Among the neutral-meson systems,
the most stringent bounds on CPT violation 
currently arise from measurements of kaon oscillations.\cite{kexpt}
Relatively little experimental information on CPT violation
is available from other neutral-meson systems.
Recently,
two groups\cite{bexpt}
at CERN have undertaken an analysis to investigate
the possibility\cite{ckpv}
that existing data can be used
to place interesting CPT bounds.
The OPAL collaboration has published the result 
$\Im\de_{B_d} = -0.020 \pm 0.016 \pm 0.006$,
while the DELPHI collaboration
has obtained a preliminary measurement 
$\Im\de_{B_d} = -0.011 \pm 0.017 \pm 0.005$.
Other analyses are underway.
Experiments of this type are important
because it is possible that relatively large CPT violation
exists for the heavier neutral mesons.
If,
for example,
the magnitude of the coefficients $a^q_\mu$ 
increases with the mass of the quark $q$,
as occurs with the conventional Yukawa couplings,
then the largest signal of CPT violation
might be expected in one of the $B$ systems.

\subsection{QED Tests}

Next,
consider some effects arising in the 
fermion sector of the extended QED.
A major type of precision test
involves the use of a Penning trap
to confine individual particles for relatively
long periods of time,
during which high-accuracy measurements of,
for instance,
anomaly and cyclotron frequencies are made.\cite{pennexpts}
The implications for these experiments
of the CPT- and Lorentz-violating terms
in the standard-model and QED extensions
have been studied.\cite{bkr}
Various signals 
and corresponding relevant figures of merit 
have been examined,
and the Lorentz and CPT reach 
attainable in present and future experiments
has been estimated.
For instance,
in the laboratory frame 
the spatial components of the coefficient $b_\mu$
are tightly constrained by experiments
comparing electron and positron anomalous magnetic moments.
The corresponding figure of merit
for CPT and Lorentz violation
could in principle be bounded 
with existing technology 
and a relatively minor change in experimental procedures
to the level of roughly $10^{-20}$.
Similarly tight bounds
on other coefficients in the fermion sector
of the extended QED 
could be obtained from other experiments of this type. 

Consider next the photon sector of the QED extension.
It turns out that the CPT-even term  
appearing in Eq.\ \rf{c}
maintains a positive conserved energy.\cite{cksm}
In contrast,
the CPT-odd term in Eq.\ \rf{b} 
can contribute negatively to the energy,\cite{cfj}
which represents a potential theoretical difficulty
and suggests the coefficient
$(k_{AF})^\ka$ should be zero.\cite{cksm}
Moreover,
it can be shown that the generalized Maxwell equations 
incorporating Lorentz violation
describe two independent propagating
degrees of freedom as usual,
but that typically the two corresponding dispersion
relations differ.
The consequences of this include
birefringence of the vacuum,
with the Lorentz violation inducing effects
on an electromagnetic wave
similar to those displayed 
by conventional radiation propagating
in an optically anisotropic and gyrotropic transparent crystal 
exhibiting spatial dispersion of the axes.

Various bounds on photon properties exist
from terrestrial, astrophysical and cosmological experiments.
Interesting constraints can be placed on the
coefficients of the Lorentz-violating terms
in the photon sector of the extended QED 
from the absence of birefringence 
on cosmological distance scales.
At present,
the components of the CPT-odd coefficient $(k_{AF})_\mu$ 
are restricted to $\lsim 10^{-42}$ GeV,\cite{cfj,hpk}
although there is a disputed claim\cite{nr,misc}
for a nonzero effect at the level of 
$|\vec k_{AF}|\sim 10^{-41}$ GeV.
Also,
the rotation-invariant irreducible component of 
the CPT-even coefficient $(k_F)_{\ka\la\mu\nu}$ 
is constrained to $\lsim 10^{-23}$ 
by the existence of cosmic rays\cite{cg}
and other tests.
The remaining irreducible components
of $(k_F)_{\ka\la\mu\nu}$ 
break rotation invariance
and could in principle be experimentally bounded 
with existing techniques
from measurements of cosmological birefringence,\cite{cksm}
although this has not yet been done.
The attainable bound on the magnitude of
the dimensionless coefficient $(k_F)_{\ka\la\mu\nu}$ 
is estimated to be of order $10^{-27}$.

The stringent experimental bound 
on the components of $(k_{AF})_\mu$ 
is compatible with the zero value
needed to minimize theoretical difficulties.
The vanishing of $(k_{AF})_\mu$ 
imposes an interesting consistency check 
on the standard-model extension because
one might expect $(k_{AF})_\mu$ 
to acquire radiative corrections 
from diagrams involving the CPT-violating couplings
in the fermion sector,
even if it is zero at tree level.
However,
it has been shown\cite{cksm}
that one-loop radiative corrections are finite,
so for one-loop renormalizability
a tree-level CPT-odd term is unnecessary. 
Similar arguments may apply at higher-loop order,
and it appears plausible to neglect 
the possible CPT-odd term in the photon sector.

In contrast,
explicit calculations have shown\cite{cksm}
that divergent radiative corrections
to the CPT-even coefficients $(k_F)_{\ka\la\mu\nu}$ 
do indeed arise at one loop,
as might be expected. 
This is compatible with 
a nonzero renormalized coupling
at the level accessible to feasible experiments
on cosmological birefringence.

\section*{Acknowledgments}
I thank Orfeu Bertolami, Robert Bluhm, Don Colladay, 
Rob Potting, Neil Russell, Stuart Samuel, 
and Rick Van Kooten for collaborations.
This work was supported in part
by the United States Department of Energy 
under grant number DE-FG02-91ER40661.

\end{document}